\documentclass[a4paper]{jpconf}
\usepackage{graphicx}
\begin{document}
\title{Third sound measurements of superfluid $^4$He films on multiwall carbon nanotubes below 1 K}

\author{Emin Menachekanian, John B.S.\,Abraham, Bob Chen, Vito Iaia, Andrew Li, and Gary A.\,Williams}
\address{Department of Physics and Astronomy, University of California, Los Angeles, CA 90095}

\ead{gaw@ucla.edu}

\begin{abstract}
Third sound is studied for superfluid films of 4He adsorbed on multiwall carbon nanotubes packed into an annular resonator. The third sound is generated with mechanical oscillation of the cell, and detected with carbon bolometers. A filling curve at temperatures near 250 mK shows oscillations in the third sound velocity, with maxima at the completion of the 4th and 5th atomic layers. Sharp changes in the Q factor of the third sound are found at partial layer fillings. Temperature sweeps at a number of fill points show strong broadening effects on the Kosterlitz-Thouless (KT) transition, and rapidly increasing dissipation, in qualitative agreement with the predictions of Machta and Guyer. At the 4th layer completion there is a sudden reduction of the transition temperature $T_{KT}$, and then a recovery back to linear variation with temperature, although the slope is considerably smaller than the KT prediction. \end{abstract}

We have carried out third sound measurements on multiwall carbon nanotubes at temperatures below 1 K, the same tubes used in the 1.3 K measurements of Ref.\,\cite{emin}.  The cell geometry is a little different than those measurements, the tubes (0.75 g) are now lightly packed in a single step into an annular resonator of the type we have used in the past \cite{kotsubo}, with mean diameter 2.9 cm and thickness 3.2 mm.  Fig.\,1 shows SEM and TEM photos of the nanotubes, which appear to 
range in diameter between 10 and 25 nm (larger than the 8-15 nm specified by the supplier \cite{tubes}) and lengths between 1-2 $\mu$m. A superconducting coil and permanent magnet assembly is attached to the resonator using thermally insulating thin-wall stainless steel tubes to mechanically vibrate the resonator, exciting the third sound.  The resonant modes are detected from the thermal oscillations by current-biased resistance bolometers, whose output goes into a LabView FFT. 
\begin{figure}[t]
\begin{center}\leavevmode
\includegraphics[width=0.9\linewidth]{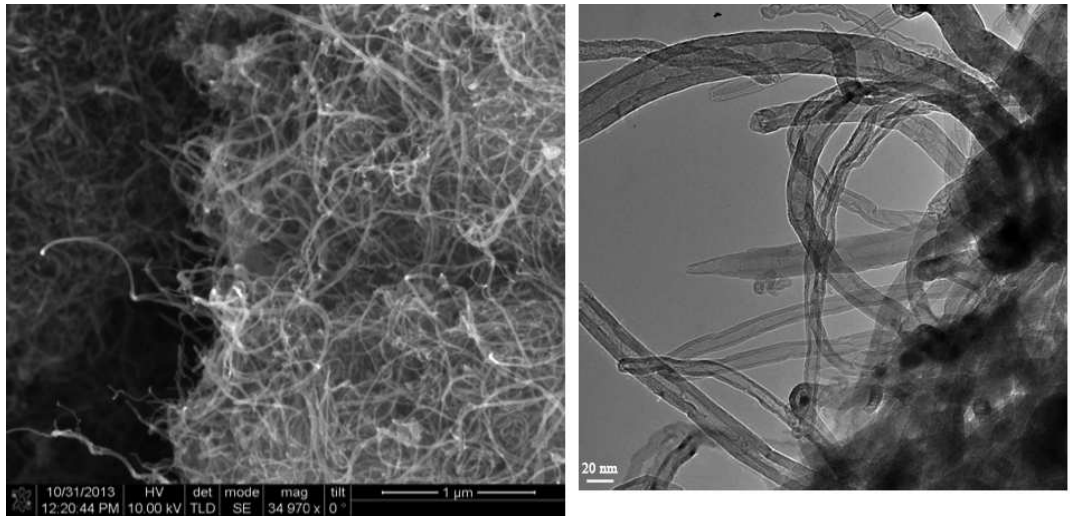}
\caption{SEM (right) and TEM (left) pictures of the multiwall carbon nanotubes}
\label{fig1}\end{center}\end{figure}
\begin{figure}[h]
\begin{center}\leavevmode
\includegraphics[width=0.9\linewidth]{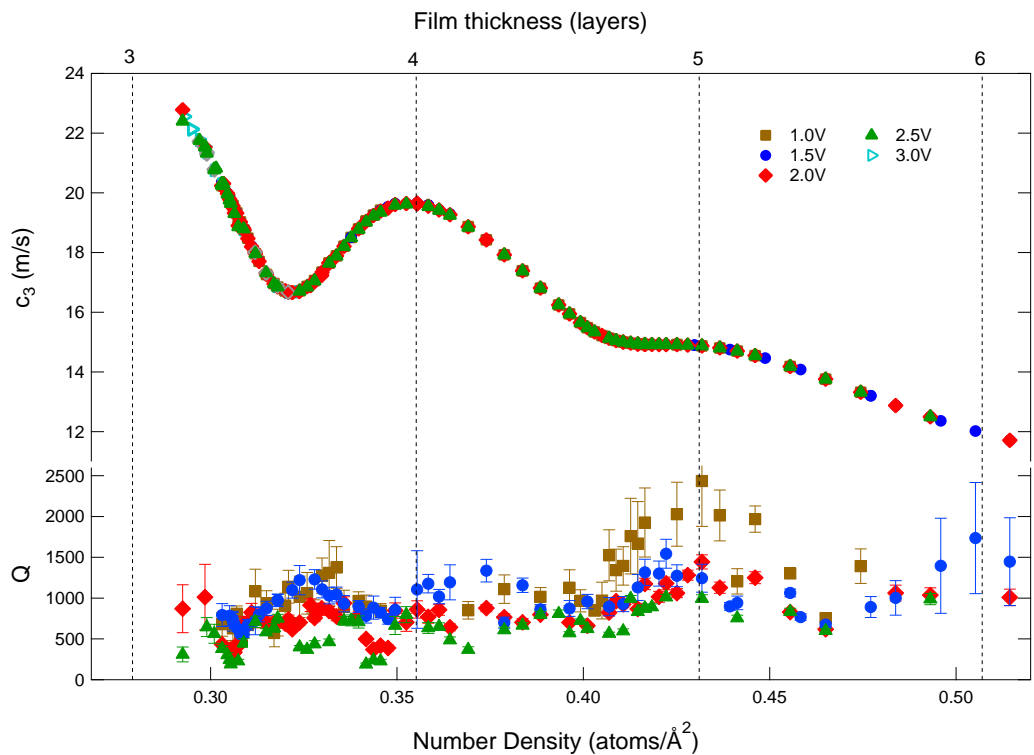}
\caption{Third sound speed and Q near 250 mK, as a function of the helium coverage.}
\label{fig1}\end{center}\end{figure}

Fig.\,2 shows the third sound speed and Q factor as a function of the helium coverage at temperatures near 250 mK.  Strong oscillations in the third sound speed are seen, due to the completion of atomic layers, quite similar to those seen in measurements on a highly oriented pyrolitic graphite (HOPG) substrate \cite{zimmerli}.  We adopt the coverage scale of those authors, taking one layer to be the relative maxima of $c_3$ at the completion of the fourth and fifth layers, and also using their determination of 0.076 atoms/\AA$^2$ as the layer coverage above the first two (more dense) layers.  Although oscillations in $c_3$ could not be detected at higher coverages, the observation of maxima in the Q factor at the 5 and 6 layer completions shows this to be a  reasonable scale.  We were completely unable to detect any signal (due to high dissipation) below 3.18 layers at oscillation levels up to 3.0 V applied to our drive coil, about the maximum we could sustain before heating of our dilution refrigerator became a problem.  The magnitude of the third sound speed is reduced from that in Ref.\,\cite{zimmerli} due to multiple scattering effects from our nanotube tangle; we deduce an index of refraction of about 1.7 comparing to their results.  The speed is only very slightly shifted down with increasing drive, while the Q is more significantly lowered (and has quite a bit of scatter).

\begin{figure}[t]
\begin{minipage}{18pc}
\includegraphics[width=19pc]{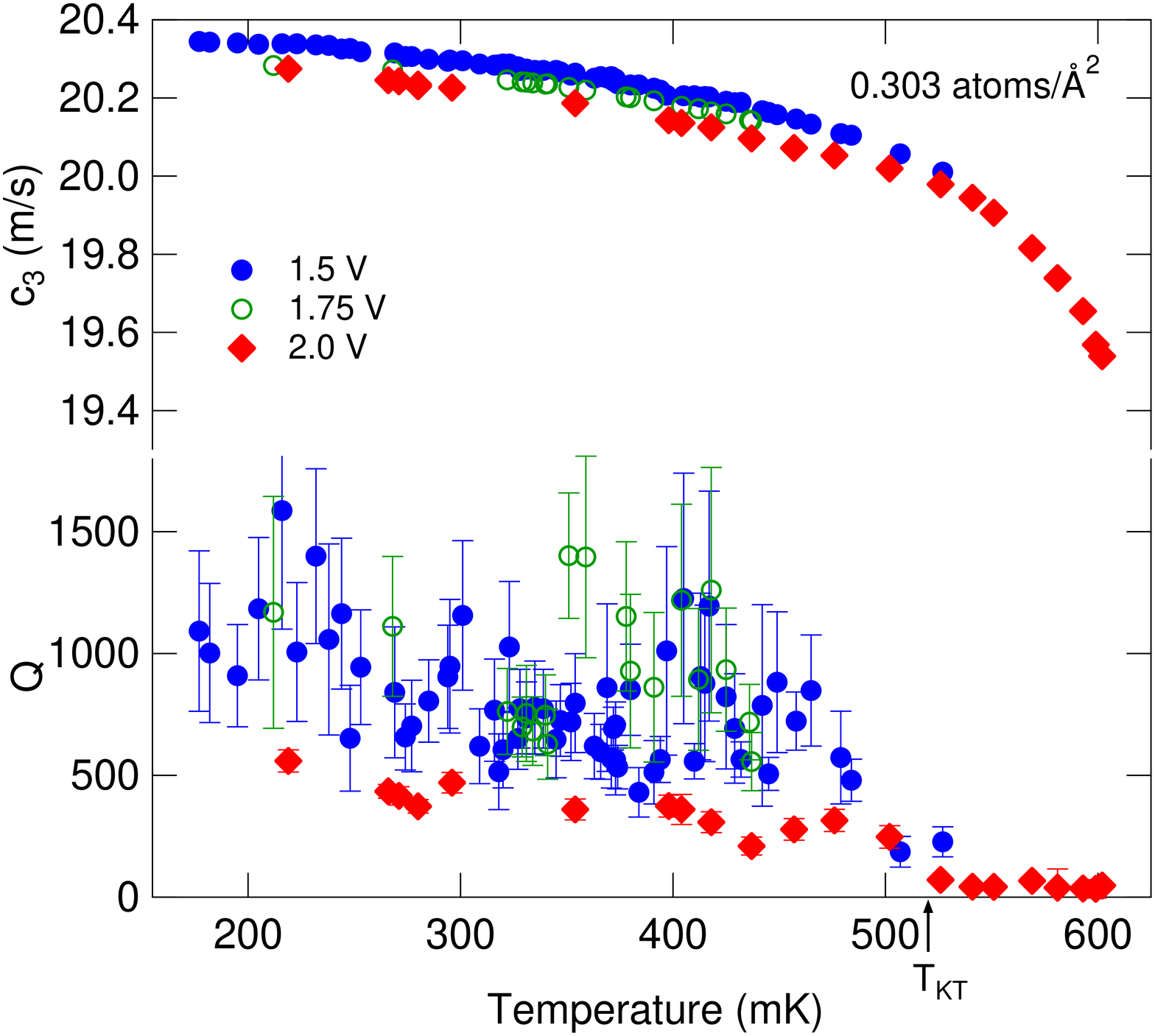}
\caption{\label{label}Temperature sweep at 3.32 layers.}
\end{minipage}\hspace{1pc}%
\begin{minipage}{19pc}.
\includegraphics[width=19pc]{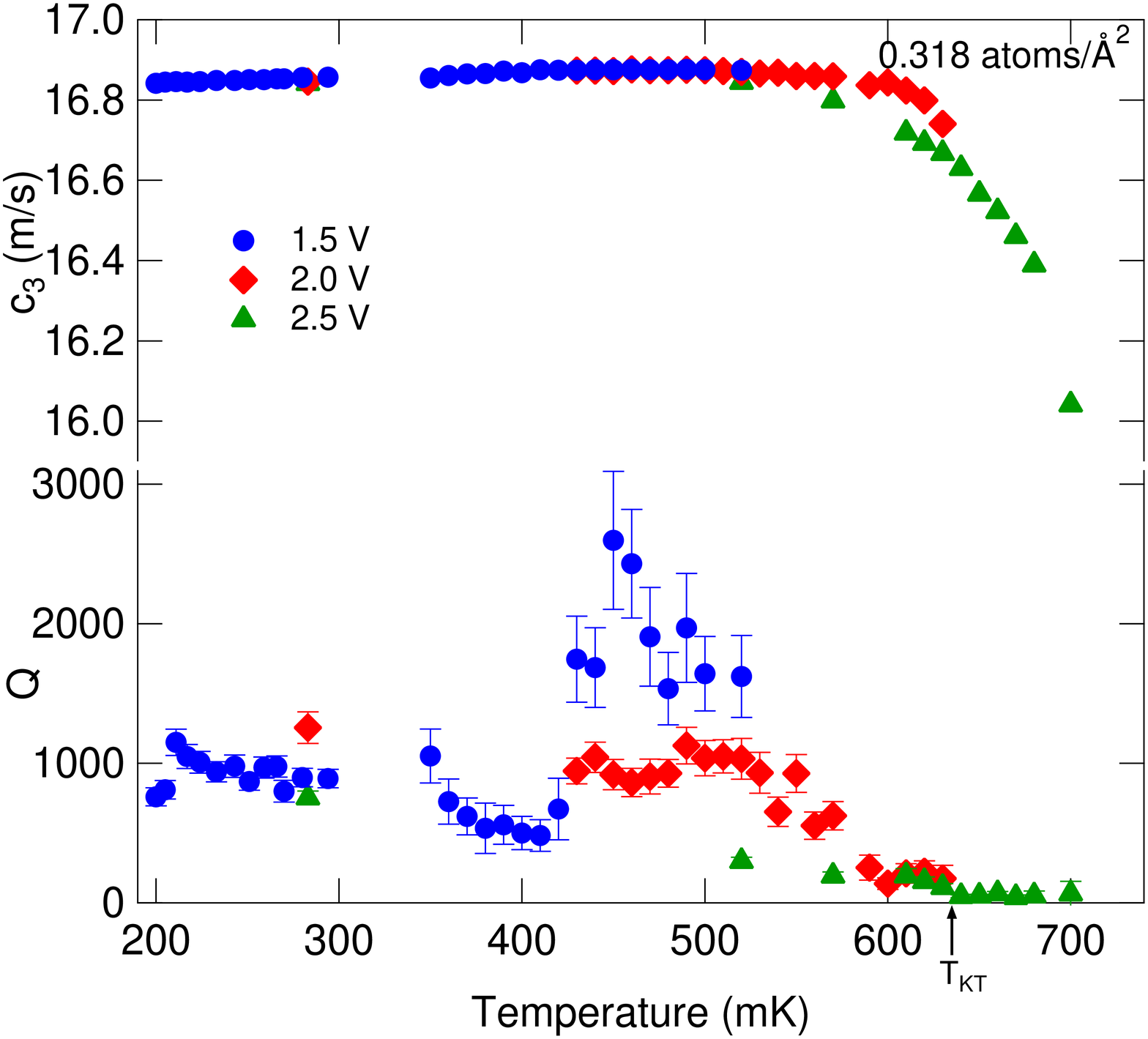}
\caption{\label{label}Temperature sweep at 3.51 layers.}
\end{minipage} 
\end{figure} 
\begin{figure}[b]
\begin{minipage}{18pc}
\includegraphics[width=19pc]{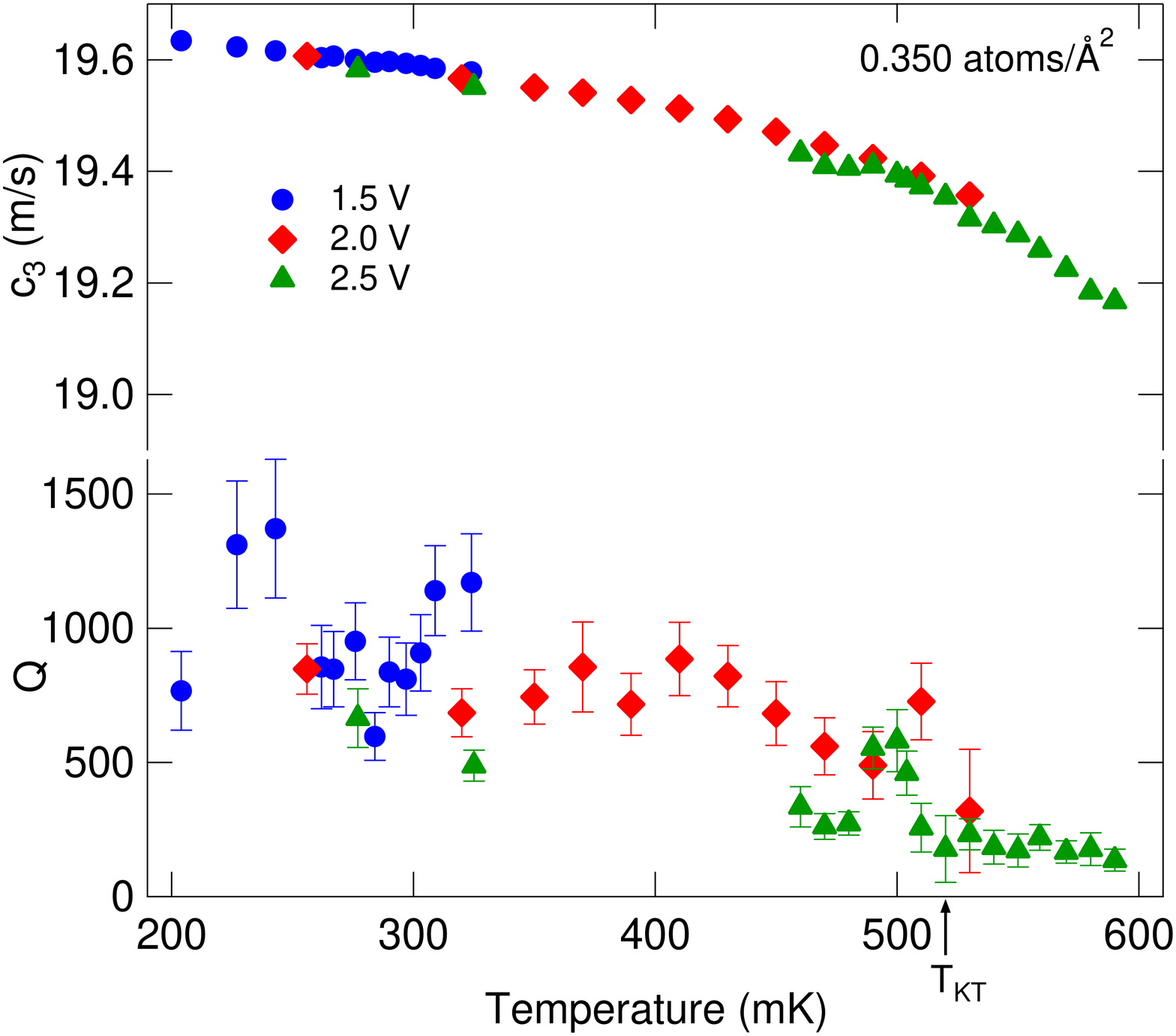}
\caption{\label{label}Temperature sweep at 3.93 layers.}
\end{minipage}\hspace{1pc}%
\begin{minipage}{18pc}
\includegraphics[width=19pc]{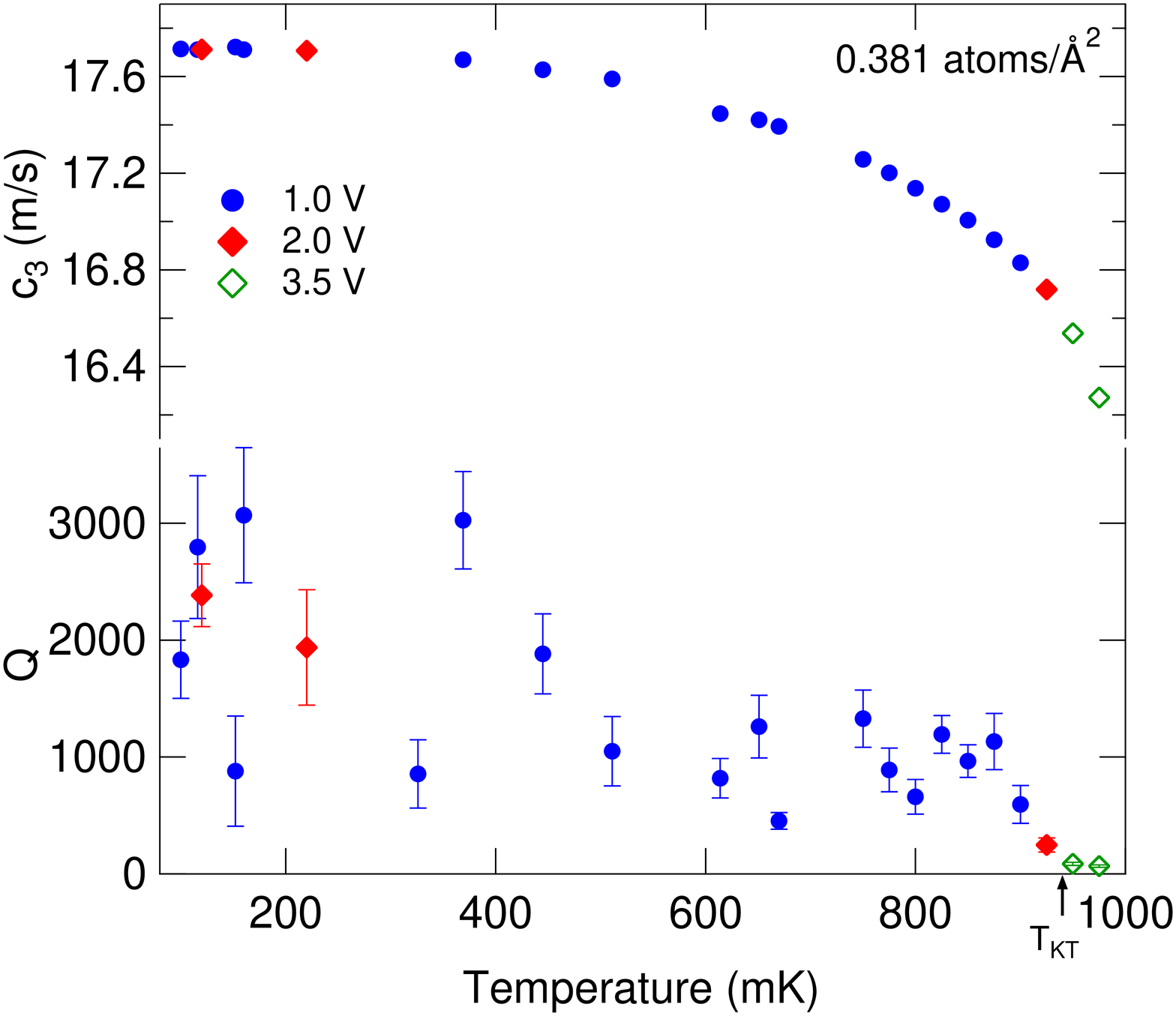}
\caption{\label{label}Temperature sweep at 4.34 layers.}
\end{minipage} 
\end{figure}

The Q is initially fairly low when the signal is first observed at 3.18 layers, and then immediately increases to a maximum before dropping rapidly to a minimum at 3.37 layers.  The origin of this dissipation is unclear, since at this low temperature there are very few thermal excitations or vortices.  The formation of a gas-liquid interface at the low-density fill of the start of the fourth layer is a possible source of this attenuation, since evaporation due to the thermal component of the third sound would be dissipative.  Such interfaces have been predicted for the second and higher layers of coverage on flat graphite substrates \cite{clements}, though it is not entirely clear if this has been observed or not \cite{crowell,greywall}.  The sharply curved nanotube substrate differs from the flat substrate, however, and nanotube simulations are only available for single-wall nanotubes and very low coverages \cite{boronat}.  With increasing coverage we find then a climb to a maximum Q near 3.5 layers, the minimum of $c_3$, showing that the high compressibility of the film at that point also plays a role in the Q factor. 

\begin{figure}[t]
\begin{center}\leavevmode
\includegraphics[width=1.0\linewidth]{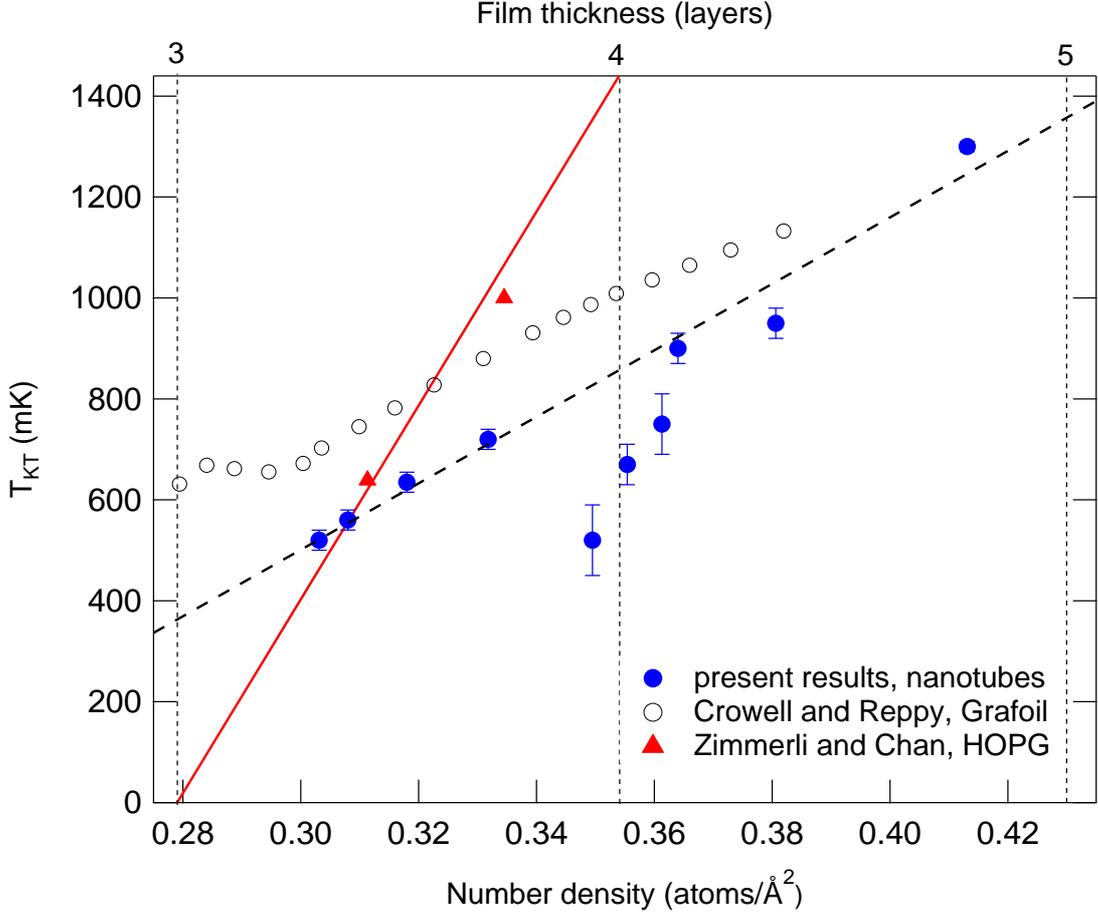}
\caption{Superfluid onset temperatures versus coverage, compared with Refs.\,\cite{zimmerli,crowell}.}
\label{fig1}\end{center}\end{figure}

Fig.\,3 shows a temperature sweep carried out at the coverage of 3.32 layers.  At low temperatures there is a slow decrease of $c_3$, with a polynomial fit showing primarily linear and square components in the decrease.  Above 520 mK there is a sharp decrease in the Q, and the beginning of a faster drop in $c_3$, and we identify this as the onset of the vortex-pair Kosterlitz-Thouless transition.  The very high dissipation prevents following $c_3$ to lower values, which is quite different behavior from our previous studies of third sound on alumina powder \cite{kotsubo} where we could follow the broadened $c_3$ to as low as 20\% of its initial value.  This observation of high dissipation on the cylindrical nanotubes, however, is in qualitative agreement with the predictions of Guyer and Machta \cite{machta} for the KT transition on cylinders, where the driven vortex pairs can counter-rotate around the cylinder, giving attenuation comparable to that on flat substrates.  They also predict a strong finite-size broadening of the transition due to the finite ratio of the nanotube radius to the vortex core radius, which is apparent from the relatively slow decrease of the third sound speed above $T_{KT}$, though due to the loss of signal this cannot be quantified.  

Fig. 4 shows a temperature sweep at 3.51 layers, close to the minimum in $c_3$.  The unusual feature here is the linear increase of $c_3$ with temperature.  This can only occur if the superfluid density increases with temperature, or if the thickness of the film decreases with temperature.  The latter possibility seems more likely, since the compressibility of the film is near its maximum here, and might further change with $T$.  However, it is also possible that internal structural changes in the film could lead to an increase in the superfluid density.  With this thicker film there is an increase in the onset temperature $T_{KT}$.  At coverages just above the minimum in $c_3$ the low-temperature speed goes back to uniformly decreasing with $T$, but now the decrease is entirely linear at the lowest temperatures.

For coverages near the maximum in $c_3$ at the fourth layer completion, we find a marked depression in the superfluid onset temperatures, as shown in Fig.\,5 at 3.93 layers, and then a further increase with higher coverage as in Fig.\,6 at 4.34 layers.  This is re-entrant superfluidity, as shown in the onset temperatures plotted in Fig.\,7:  adding coverage at a fixed temperature of say 600 mK would see superfluidity disappear at 3.8 layers, and then reappear at 4.1 layers.  This behavior at the fourth layer completion appears to be unique to the nanotube films, as it is not observed on the Grafoil  substrate \cite{crowell}, whose onset temperatures are also plotted on the same graph.  The Graphoil data only shows a relatively constant $T_{KT}$ near the third layer completion (visible in Fig.\,7), and nothing similar at the fourth layer completion.  There have been simulations \cite{zimanyi} predicting a drop in superfluid density at layer completions, from the increasing atomic repulsion, which would translate to a drop in $T_{KT}$ such as we observe.  The effect seems to be much stronger on the nanotubes compared to flat substrates.

The dashed curve in Fig.\,7 shows a linear fit to our onset temperatures (excluding the three points near the fourth layer completion), with an onset thickness extrapolating to 2.4 layers, similar to that found in \cite{crowell} in the third layer. The slope of this curve, however, is only about 1/2 of the expected KT universal value, shown as the solid line in the plot for an onset thickness of three atomic layers as deduced on the HOPG substrate \cite{zimmerli}. This is unusual, but it also seems to be the same result found on the Grafoil substrate \cite{crowell}, where we have plotted the reported dissipation maximum, which should track $T_{KT}$. It appears as if only about 1/2 of the atoms that are being added actually join the superfluid condensate. It is unclear why this occurs: possibilities are that there could be a strong localization of atoms at defects on the substrate, or if gas-liquid phase separation occurs, the gas atoms might not be superfluid, as speculated in \cite{crowell}. The HOPG substrate experiment, however, appeared to find results consistent with the universal KT line, but unfortunately with only two data points. Our finding in Fig. 2 that third sound could not be observed at any temperature below about 3.2 layers brings up the question of whether the onset seen in Ref.\,\cite{zimmerli} at 3.4 layers and 0.639 K was an actual KT onset, or simply the same third sound signal loss from high dissipation we observed on approaching the third-layer completion point.

We have found no evidence in our measurements of helium adsorption inside the nanotubes, though that may have occurred. The TEM pictures seem to show a few tubes with what appear to be open ends, but also some of the tubes show closed end caps.  If helium did enter it would have been at the very beginning of the fill, due to the strong effects of surface tension.

\section*{Acknowledgements}

We thank William Hubbard and Chris Regan for taking the electron microscope photos at the UCLA NanoSystems Institute.  This work was supported in part by the U.\,S.\, National Science Foundation, Grant No. DMR 0906467.

\section*{References}
\bibliography{Nanotube}
\end{document}